# Emergence of Competing Stripe Phase near the Mott Transition in Ti-doped Bilayer Calcium Ruthenates

Ashish Gangshettiwar[1†], Yanglin Zhu[2†], Zhanzhi Jiang[1], Jin Peng[3], Yu Wang[2,3], Jiaming He[4], Jianshi Zhou[4], Zhiqiang Mao[2,3], Keji Lai[1*]

[1]Department of Physics, University of Texas at Austin, Austin, Texas 78712
[2]Department of Physics, The Pennsylvania State University, University Park, Pennsylvania 16802-6300
[3]Department of Physics, Tulane University, New Orleans, Louisiana 70118
[4]Department of Mechanical Engineering, University of Texas at Austin, Austin, Texas 78712

[†]Equal contributions to the paper.
[*]Email: kejilai@physics.utexas.edu

## Abstract

We report the nanoscale imaging of Ti-doped bilayer calcium ruthenates during the Mott metal-insulator transition by microwave impedance microscopy. Different from a typical first-order phase transition where coexistence of the two terminal phases takes place, a new metallic stripe phase oriented along the in-plane crystalline axes emerges inside both the G-type antiferromagnetic insulating state and paramagnetic metallic state. The effect of this electronic state can be observed in macroscopic measurements, allowing us to construct a phase diagram that takes into account the energetically competing phases. Our work provides a model approach to correlate the macroscopic properties and mesoscopic phase separation in complex oxide materials.

The Ruddlesden–Popper series of alkaline-earth ruthenates $(Sr,Ca)_{n+1}Ru_nO_{3n+1}$ display a wealth of fascinating behaviors that are representative of strongly correlated systems [1]. Compared with the Sr-based compounds [2, 3], the metallicity of the Ca-based counterparts, if any, is much weaker due to the more distorted crystal structures [1]. For instance, $CaRuO_3$ ($n = \infty$) is a paramagnetic 'bad' metal (PM-M) close to the antiferromagnetic (AFM) instability [4], whereas the single-layer ($n = 1$) $Ca_2RuO_4$ is a G-type AFM Mott insulator (G-AFM-I) at room temperature and undergoes a metal-insulator transition (MIT) at $T_{MIT} = 357$ K [5]. The intermediate member of bilayer ($n = 2$) $Ca_3Ru_2O_7$ is more complex, showing a magnetic transition at $T_N = 56$ K and an MIT at $T_{MIT} = 48$ K [6 – 10]. The magnetic ordering below $T_N$ is of A-type AFM, i.e., ferromagnetic bilayers stacked antiferromagnetically along the $c$ axis, and the moments switch from the $a$ axis (denoted as AFM-$a$) to the $b$ axis (AFM-$b$) upon cooling, as illustrated in Fig. 1a [10]. Moreover, the small Fermi pocket in the AFM-$a/b$ phases [11] is suppressed by isovalent Ti doping into the Ru site through bandwidth reduction, which drives the ground state of $Ca_3(Ru_{1-x}Ti_x)_2O_7$ into G-AFM-I beyond $x = 3\%$ [12 – 17]. Such simultaneously active lattice, charge, and spin degrees of freedom make Ti-doped $Ca_3Ru_2O_7$ an ideal testbed to explore the correlation physics in complex oxides.

The multiple phases in $Ca_3(Ru_{1-x}Ti_x)_2O_7$ with distinct electrical and magnetic properties (Fig. 1a) are analogous to the colossal magnetoresistive (CMR) manganites [18]. As in any first-order phase transitions, the MIT in most CMR systems is accompanied by a mixture of two terminal phases, which underlies the drastic change of resistivity under external stimuli such as temperature ($T$), magnetic ($B$) field, electric current, light, and pressure [19]. In this Letter, we report the direct visualization of coexisting phases across the Mott transition in 10% Ti-doped bilayer calcium ruthenates using near-field microwave microscopy. Surprisingly, within a narrow range of $T$ and $B$-field near the transition, stripe-like metallic domains oriented along the $a$-axis, which differ from the two terminal phases, appear inside both the G-AFM-I and PM-M regions. Based on the dynamic emergence of the mesoscopic phases and fine features in the macroscopic transport and magnetization data, a phase diagram that includes the phase coexistence can be constructed. Our observation of orientation-ordered phase separation suggests that strongly correlated materials with $4d$ electrons share certain common aspects, while differ in others, with the $3d$ correlated electron systems.

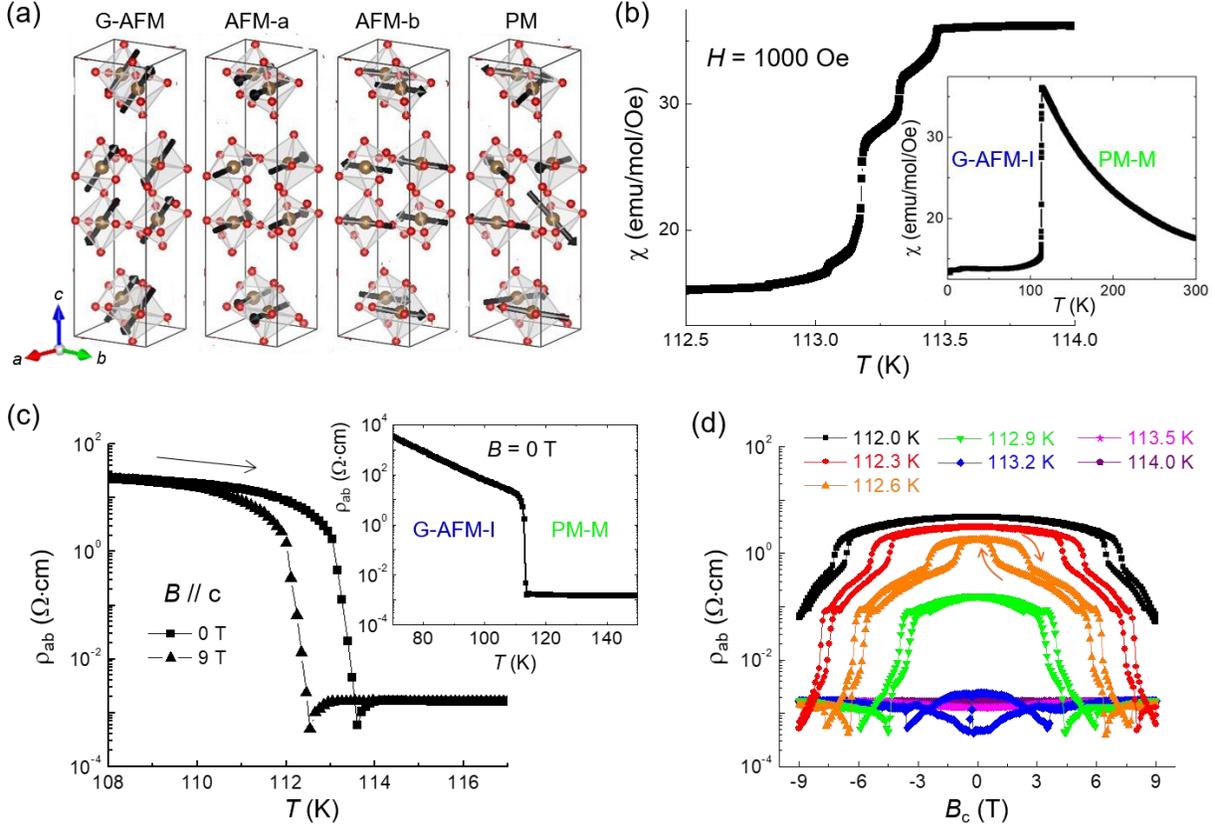

FIG. 1. (a) Schematics of the magnetic structures of (from left to right) G-AFM, AFM-*a*, AFM-*b*, and PM phases [10, 12]. Only the RuO$_6$ octahedrons are shown here for clarity. (b) Magnetic susceptibility taken under an out-of-plane field of 1000 Oe, showing multiple meta-magnetic steps around the transition. The inset shows $\chi(T)$ in a wider temperature range and coarse steps, where only a single transition is observed. (c) In-plane resistivity taken at $B_c = 0$ T and 9 T. The inset shows $\rho_{ab}(T)$ at 0 T in a wider temperature range and coarse steps. (d) Magnetoresistivity at various temperatures. The field sweep direction is indicated for the 112.6 K data.

Single crystals Ca$_3$(Ru$_{0.9}$Ti$_{0.1}$)$_2$O$_7$ in this study were grown by the floating zone technique [12]. The *T*-dependent magnetic susceptibility ($\chi$) and in-plane resistivity ($\rho_{ab}$) data with relatively coarse steps (0.05 K for $\chi$ and 0.5 K for $\rho_{ab}$) are plotted in the insets of Figs. 1b and 1c, respectively, showing a single transition from G-AFM-I to PM-M at ~ 113 K upon warming. The results appear to agree with prior investigations on the same $x = 10\%$ material, where no intermediate states between the two phases were found [13, 15]. The situation, however, was different when much finer measurements were taken (0.01 K step for $\chi$ and 0.1 K for $\rho_{ab}$), as shown in Figs. 1b and 1c. In the $\chi(T)$ curve, several jumps characteristic for meta-magnetic transitions were observed. Within a narrow window in the resistivity data, $\rho_{ab}(T)$ is lower than that of the PM-M phase, reminiscent of the parent compound where the metallic AFM-*a/b* phase develops [6, 10]. Figs. 1c and 1d also show the result under an out-of-plane *B*-field, which is compatible with our scanning experiment. It should be noted that the spin flip-flop transition for *B* // *c*-axis occurs at much higher fields than *B* // *a*- or *b*-axis and intermediate states are present throughout the transition [20]. As a result, the effect of $B_c$ in this work is mostly to destroy the G-AFM ordering. The presence of multiple steps and sudden jumps in the magnetoresistance data indicates that the MIT may involve richer physics than previously conjectured. In particular, the resistivity at 113.2 K jumps between $2 \times 10^{-3}$ Ω·cm (PM-M) and a lower value of $4 \times 10^{-4}$ Ω·cm, indicative of the emergence of a new metallic phase near the transition.

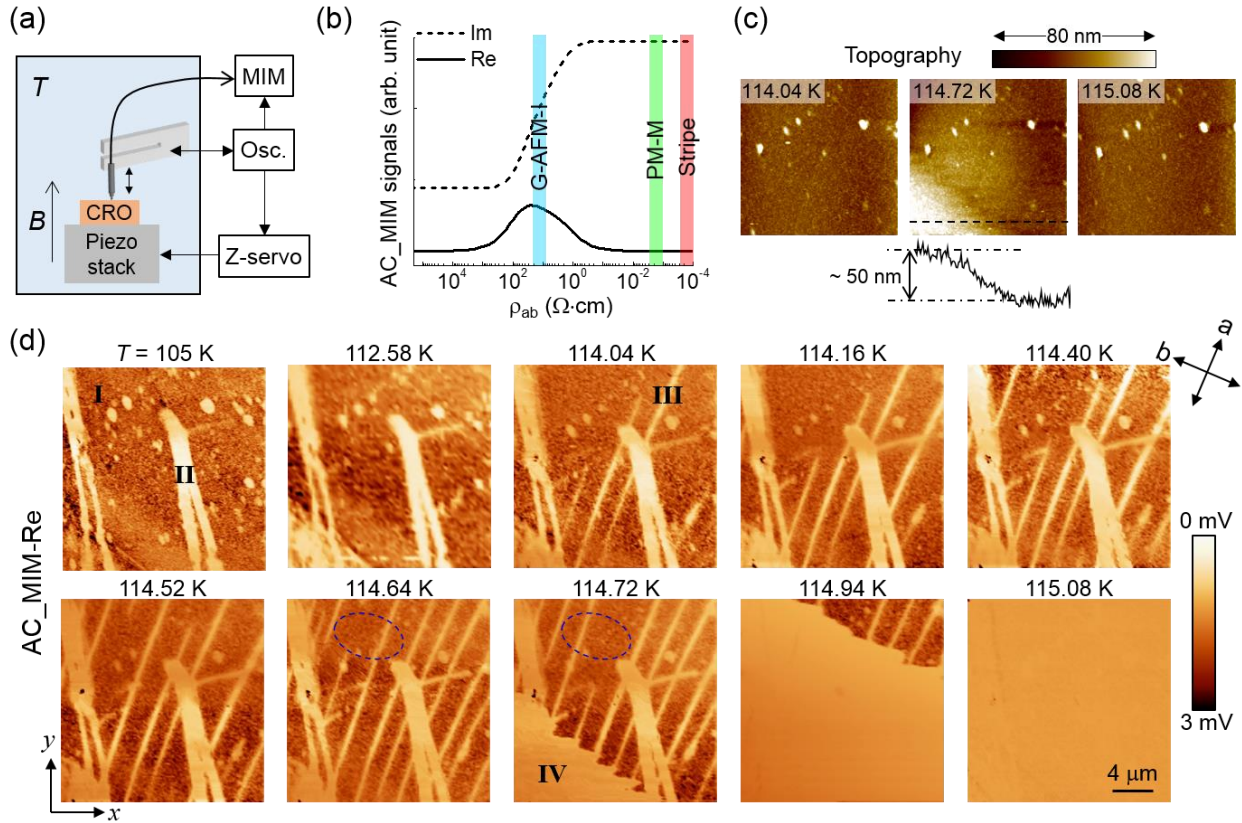

FIG. 2. (a) Schematic of the tuning-fork-based cryogenic MIM setup. The TF electronics control the z-scanner and provide the reference to demodulate the MIM signals. (b) Simulated AC_MIM signals as a function of $\rho_{ab}$. The resistivity values of relevant phases are indicated in the plot. (c) Topographic images at three selected temperatures. Note that the color scale corresponds to relative height difference within each frame. The inset shows a line profile in the 114.72 K image. (d) AC_MIM-Re images at various temperatures across the transition. Phases I to IV are labeled in the images. The blue ellipses in the 114.64 K and 114.72 K images show the dynamic appearance and disappearance of two stripes. The scanning directions (*x* and *y*) and the crystalline axes (*a* and *b*) are indicated in the bottom left and top right corners, respectively. All images are 20 μm × 20 μm.

In order to explore the real-space evolution of this complex phase transition, we carried out cryogenic microwave impedance microscopy (MIM) [21], as illustrated in Fig. 2a. The 1 GHz signal is delivered to a tungsten tip (diameter ~ 100 nm) glued to a quartz tuning fork (TF) for topographic feedback [22, 23]. The microwave electronics measure the real (MIM-Re) and imaginary (MIM-Im) parts of the tip-sample admittance, which is demodulated at the TF resonant frequency (~ 40 kHz) to form the corresponding AC_MIM images [23]. Fig. 2b shows the simulated AC_MIM-Im/Re signals as a function of the in-plane resistivity (Fig. S1, [24]), taking into the fact that $\rho_{ab} \ll \rho_c$ in this material [15, 16]. Even though the low-$T$ phase is insulating, its resistivity near the MIT is still on the conductive side for the 1 GHz MIM [23]. On this side of the response curve, it is easier to distinguish the PM-M and G-AFM-I regions in the MIM-Re channel, where the signal decreases as the sample becomes more conductive, due to the cleaner data and less topographic cross-talk than MIM-Im. In addition, surface particles observed in the topography (Fig. 2c) also display vanishing MIM-Re signals. Nevertheless, they are clearly on the resistive side of Fig. 2b as their MIM-Im signals are the lowest in the map (Fig. S2, [24]).

The AC_MIM-Re images at various temperatures across the MIT are shown in Fig. 2d (complete data in Fig. S2, [24]). Note that the apparent $T_{MIT}$ in Figs. 1 – 3 may differ slightly by a few Kelvins, depending on the detailed thermal or magnetic-field history of the sample, which is not uncommon for strongly correlated materials. For temperatures much below $T_{MIT}$, the sample was mostly in the insulating state (Phase I), with some highly conductive areas (Phase II) observed in the microwave image. Since $T_{MIT}$ decreases as decreasing Ti-doping, Phase II is likely associated with local deficiency of Ti concentration, which segregate into isolated regions during the floating-zone crystal growth. Since the lattice constant sensitively depends on the Ti doping, such minority phases may self-organize in regular shapes under strain when the crystal is cooled to room temperature, which could explain its ribbon-like shape and orientation along the high-symmetry [110] and [1-10] directions. Finally, we note that Phase II does not exhibit obvious topographic features throughout the phase transition (Fig. 2c). As the *c*-axis lattice constant is rather different between G-AFM-I and PM-M phases [12, 16], Phase II regions in the MIM images must be thin slabs on the surface. One cannot, however, exclude the existence of similar Ti-deficient domains in the interior of the sample.

Starting from 114.04 K, metallic stripes (Phase III) were observed in the MIM images, whose length and areal density grew rapidly with increasing $T$ at 0.1 K steps. The stripes are oriented along the *a*-axis of the crystal, as determined by the X-ray diffraction (XRD) data (Fig.

S3, [24]). The width of the stripes ranges from 100 nm (limited by the spatial resolution) to ~ 1 μm and the spacing between adjacent ones is 3 ~ 5 μm. The micrometer-sized separation between the stripes indicates that the elastic strain between different phases, rather than the electronic correlation, plays a key role here. In particular, the characteristic spacing is likely determined by the length scale at which the strain can be accommodated inside the crystal. The appearance of stripes is dynamic (blue ovals in Fig. 2d) and they are not pinned to specific locations in different thermal cycles (Fig. S4, [24]). At 114.72 K, another metallic phase (Phase IV) set in, which quickly swept through the scanned area with an additional ~ 0.3 K. The topographical images in Fig. 2c indicate that Phase IV is associated with a pronounced increase in height, corresponding to the ~ 0.1% increase of lattice constant in the $c$-axis across the MIT [12, 16]. Phase IV is thus the global PM-M phase that extends through the entire crystal.

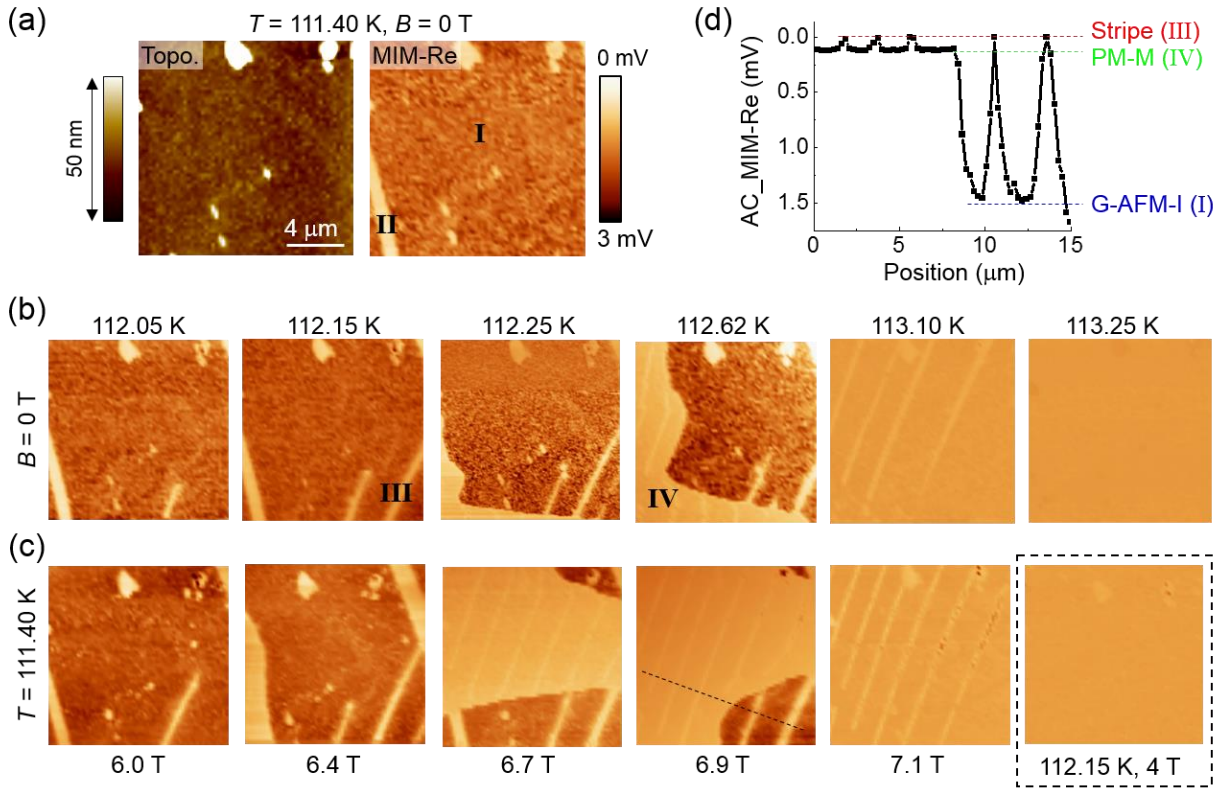

FIG. 3. (a) Topographic and AC_MIM-Re images at 111.40 K and 0 T of a different cool down. (b) MIM images at various temperatures and 0 T. (c) MIM images at various magnetic fields and 111.40 K, except for the last image at a higher temperature and 4 T. Phases I to IV are again labeled in the data. All images are 15 μm × 15 μm. (d) Line profile in (c), showing the AC_MIM-Re signals across stripes in both the G-AFM-I and PM-M phases.

Since the MIM contrast between the two possible metallic phases (PM-M and AFM-$a/b$) is too small, it is not obvious from Fig. 2 whether Phase III is the same as II/IV or represents a new electronic state. To enhance the contrast, we have used a blunter tip in a different cool down (Fig. S1, [24]). Fig. 3a shows the topography and AC_MIM-Re images at $T = 111.40$ K and $B = 0$ T, from which two rounds of experiments were performed. In Fig. 3b, the $B$-field was kept at zero during a warm-up across the transition. Similar to Fig. 2c, all four phases – I as G-AFM-I, II as local PM-M at lower $x$, III as the stripes, and IV as the global PM-M – were observed in this area. Strikingly, the stripes could also be seen inside Phase IV. The same phenomenon was also observed in Fig. 3c, where we kept $T = 111.40$ K and ramped up the $B$-field, except for the last image (Fig. S5, [24]). Moreover, the line profile in Fig. 3d shows that the stripe domains inside Phases I and IV display the same MIM signals. In other words, the stripes are indeed a new electronic state that is distinct from the two terminal phases of the transition. To our knowledge, this is the first observation of an additional phase during the metal-insulator transition in strongly correlated materials, highlighting the complexity in 4$d$-ruthenates. Interestingly, the stripes in the two major phases seem to avoid each other at the boundary (Fig. S6, [24]). Based on the prior knowledge in $Ca_3(Ru_{1-x}Ti_x)_2O_7$ [12 – 17], the stripes are likely the sequential appearance of AFM-$a$ and AFM-$b$ phases during the transition, although we do not have direct evidence on their magnetic ordering. Besides, Phase III does not exhibit appreciable topographic contrast over the G-AFM-I or PM-M phases, whose lattice constants in the $c$-axis differ by ~ 0.1% [12, 16]. Since the noise floor of our TF feedback is 2 ~ 3 nm, one can infer that the stripes in the MIM images are surface features with a thickness no more than a few microns. Similar to the analysis of Phase II, however, we cannot exclude the appearance of such domains inside the crystal (Fig. S2, [24]).

The mesoscopic MIM imaging allows us to reevaluate the magnetization and transport data and construct a new phase diagram for $Ca_3(Ru_{0.9}Ti_{0.1})_2O_7$ [13, 15, 25] that includes the nanoscale phase separation. For instance, the magnetotransport curve at 112.6 K (Fig. 4a) can be divided into four sections based on sudden changes of the slope, presumably due to the appearance or disappearance of certain coexisting phases. Similar analysis can be applied to the $\rho(T)$ and $\chi(T)$ data in Fig. 1. As shown in Fig. 4b, the coexistence of the stripe phase with the two terminal phases reveals a complex energy landscape that takes place near the Mott transition in this system.

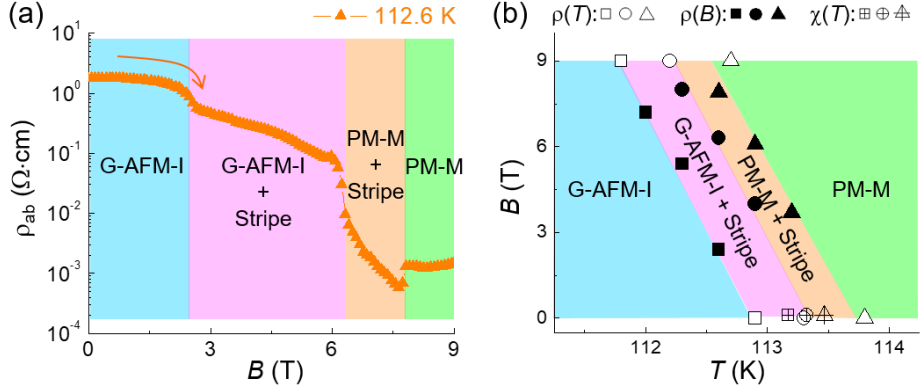

FIG. 4. (a) Magnetotransport curve at $T = 112.6$ K divided into four sections based on sudden changes of the signals. (b) Phase diagram in a narrow temperature and field range taking into account the phase separation. The squares, circles, and triangles (open, solid, crossed from $\rho(T)$, $\rho(B)$, and $\chi(T)$, respectively) mark the three phase boundaries indicated in the graph. For consistency, all data points are measured on the insulator-to-metal part of the hysteresis loop, i.e., warming up for $T$-dependence and ramping up for $B$-dependence.

As a concluding remark, it is instructive to compare our results with nanoscale phase separation in other complex oxides. In the more extensively studied $3d$ correlated electron systems near MITs, the coexisting phases may exhibit random shapes and sizes [26 – 30] or orientation-ordered patterns [31 – 34]. The latter clearly signifies the strong effect of elastic strain on phase transitions. For comparison, stripe-like competing phases are commonly seen in single crystals of $4d$ ruthenate perovskites, such as $Ca_2RuO_4$ [35], Mn-doped $Sr_3Ru_2O_7$ [36], and now $Ca_3(Ru_{1-x}Ti_x)_2O_7$. It is possible that, as the on-site Coulomb energy is reduced for the more extended $4d$ orbitals with respect to the $3d$ counterpart, the lattice degree of freedom in ruthenates becomes prominent for bulk ruthenates materials. Subtle difference is also expected between different members in the Ruddlesden–Popper series. For instance, stripe domains that are the same metallic phase at the end of the insulator-to-metal transition were also reported in $Ca_2RuO_4$ [35], which occur near the physical boundary between the two competing phases. In our case, the stripes, which differ from the two terminal states of the Mott transition, are not confined to the interface between the two global phases. In fact, the effect of phase separation is observable in macroscopic measurements. This new complexity is likely due to the fact that more energetically competing phases are available in the bilayer than single-layer ruthenates.

In summary, we report the nanoscale microwave imaging on doped $Ca_3(Ru_{1-x}Ti_x)_2O_7$ through the simultaneous insulator-to-metal and antiferromagnetic-to-paramagnetic transition. Different from the previous picture that a single-step process takes place, we observed a stripe-like metallic phase within a narrow temperature and field range of the transition, which is in a different electronic state from the two majority phases. The emergence of such orientation-ordered domains is consistent with the macroscopic transport and magnetization data, suggesting the strong interplay between electronic and lattice degrees of freedom in 4$d$ correlated electron systems.

The MIM work (A.G., Z.J., and K.L.) was supported by the U.S. Department of Energy (DOE), Office of Science, Basic Energy Sciences, under the award no. DE-SC0019025. Financial support for sample preparation was partially provided by the National Science Foundation through the Penn State 2D Crystal Consortium-Materials Innovation Platform (2DCC-MIP) under NSF cooperative agreement DMR-1539916.

[34]    M. Liu, A. J. Sternbach, M. Wagner, T. V. Slusar, T. Kong, S. L. Bud'ko, S. Kittiwatanakul, M. M. Qazilbash, A. McLeod, Z. Fei et al., *Phys. Rev. B* **91**, 245155 (2015).

[35]    J. Zhang, A. S. McLeod, Q. Han, X. Chen, H. A. Bechtel, Z. Yao, S. N. Gilbert Corder, T. Ciavatti, T. H. Tao, M. Aronson et al., *Phys. Rev. X* **9**, 011032 (2019).

[36]    T.-H. Kim, M. Angst, B. Hu, R. Jin, X.-G. Zhang, J. F. Wendelken, E. W. Plummer, and A.-P. Li, *Proc. Natl. Acad. Sci.* **107**, 5272 (2010).11

Supplementary materials to

# Emergence of Competing Stripe Phase near the Mott Transition in Ti-doped Bilayer Calcium Ruthenates

Ashish Gangshettiwar[1†], Yanglin Zhu[2†], Zhanzhi Jiang[1], Jin Peng[3], Yu Wang[2,3], Jiaming He[4], Jianshi Zhou[4], Zhiqiang Mao[2,3], Keji Lai[1*]

[1]Department of Physics, University of Texas at Austin, Austin, Texas 78712
[2]Department of Physics, The Pennsylvania State University, University Park, Pennsylvania 16802-6300
[3]Department of Physics, Tulane University, New Orleans, Louisiana 70118
[4]Department of Mechanical Engineering, University of Texas at Austin, Austin, Texas 78712

[†]Equal contributions to the paper.
[*]Email: kejilai@physics.utexas.edu

## S1. Finite-element analysis of the AC_MIM signals

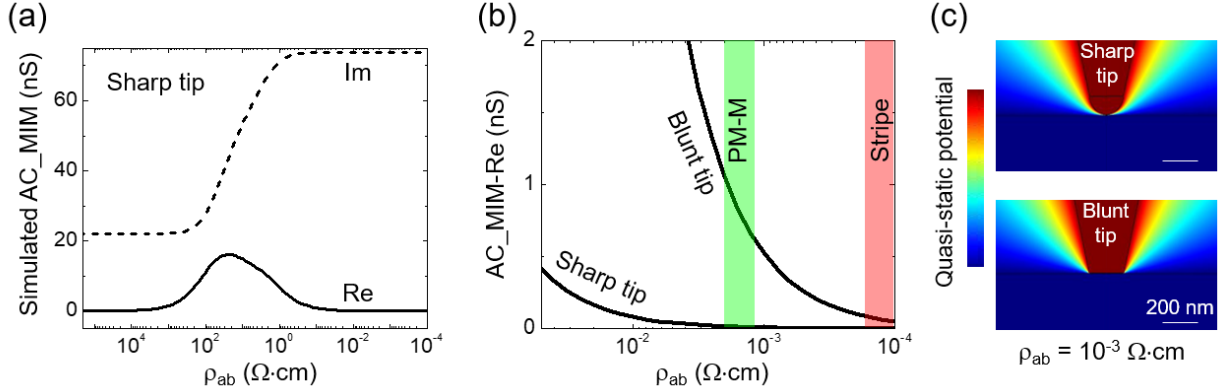

FIG. S1. (a) Simulated AC_MIM signals as a function of the in-plane resistivity for a sharp tip. (b) Simulated AC_MIM-Re signals for sharp and blunt tips. The dc resistivity values of the PM-M and stripe phases are indicated in the plot. (c) Simulated 1 GHz quasi-static potential distribution around two tips for $\rho_{ab} = 10^{-3}$ Ω·cm.

To understand the AC_MIM signals, we performed finite-element analysis (FEA) to calculate the tip-sample admittance, as shown in Fig. S1a. Here we first computed the approaching curve towards the sample surface, followed by demodulation of AC_MIM-Im/Re signals for a sinusoidal tapping of the tip (amplitude ~ 15 nm in this work). Detailed simulation procedure can be found in Ref. S1. To avoid the divergence of the signals as the tip is in contact with a metallic sample, we assume that the minimal tip-sample spacing is 1 nm. Note that $Ca_3(Ru_{1-x}Ti_x)_2O_7$ is highly anisotropic between the in-plane and out-of-plane directions [S2, S3]. We assume that $\rho_c = 100\, \rho_{ab}$ in the FEA modeling.

The MIM signals are very sensitive to the exact tip conditions in the experiment. It is possible that the tip used to take the images in Fig. 3 of the main text is blunter than that in Fig. 2. As shown in Fig. S1b, while a sharp tip (semispherical apex with 100 nm radius) cannot differentiate the PM-M and stripes, a very blunt tip (flat end with 100 nm radius) gives much higher contrast between the two phases. We acknowledge that this effect alone cannot quantitatively explain all features in the data, such as the comparable contrast between G-AFM-I and PM-M phases in Figs. 2 and 3 of the main text. Given the many approximations in the FEA modeling (1 nm minimum tip-sample spacing, $\rho_c = 100\, \rho_{ab}$, same conductivity at dc and GHz…), we believe our simulation provides a semi-quantitative framework for the data interpretation. The conclusion that three phases (G-AFM-I, stripe, and PM-M) are observed during the Mott transition is not affected by the simulation result.

## S2. Complete set of *T*-dependent data

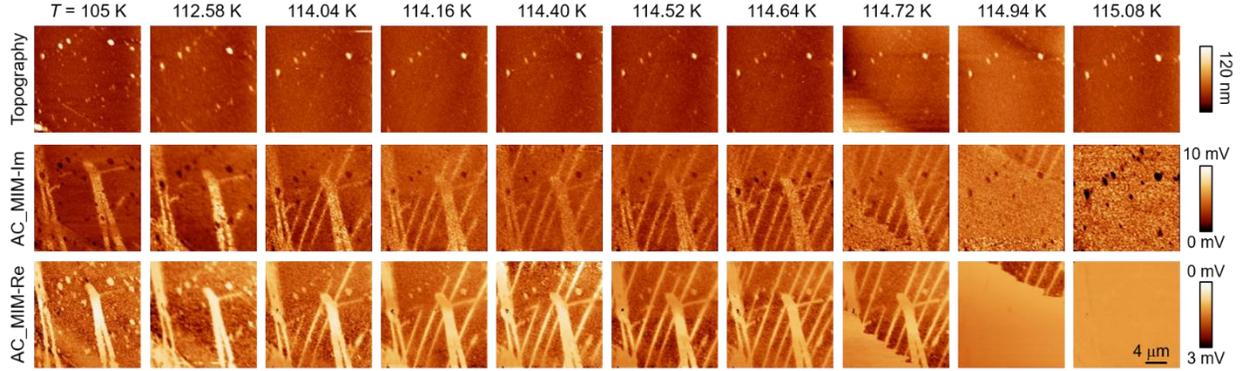

FIG. S2. Complete set of the *T*-dependent topography and AC_MIM-Im/Re images.

The complete topography and AC_MIM images in Fig. 2 of the main text are shown in Fig. S2. The signals in the Re channel are generally cleaner due to the higher contrast and less topographic cross-talk. Since AC_MIM-Re signals on the conductive side of the response curve (see Fig. S1) are decreasing as increasing sample conductivity, we flip the color scale in the data. The surface particles appear as dark spots in AC_MIM-Im, suggesting that they are not conductive.

In the MIM images, one can also observe a few conductive stripe-like features along the [1-10] direction well below the Mott transition temperature, which are more pronounced in the MIM-Im channel than the MIM-Re data. Similar behaviors are seen in other areas during our experiment. In all, the slab-like features along the [110] and [1-10] directions are Ti-deficient domains (Phase II) and the stripe-like features along the *a* axis are the emergent AFM-*a/b* phase (Phase III).

Similar analysis also applies for Phase III. In general, the resistivity measured in transport experiments is a bulk property. However, the bilayer CaRuO-327 system is strongly quasi-2D with $\rho_c/\rho_{ab} \sim 10^3$ near the Mott transition. In other words, for the sample with a thickness of 100 ~ 200 μm in this study (see Fig. S3a for a picture of the crystal), a surface layer of 1 μm in height could dominate the in-plane transport. We nevertheless recognize that the MIM only has a probing depth comparable to the tip diameter, i.e., on the order of 100 nm. As a result, one cannot exclude the possibility that such AFM-*a/b* domains also occur in the interior of the crystal.

## S3. Determine crystal orientation by X-ray diffraction

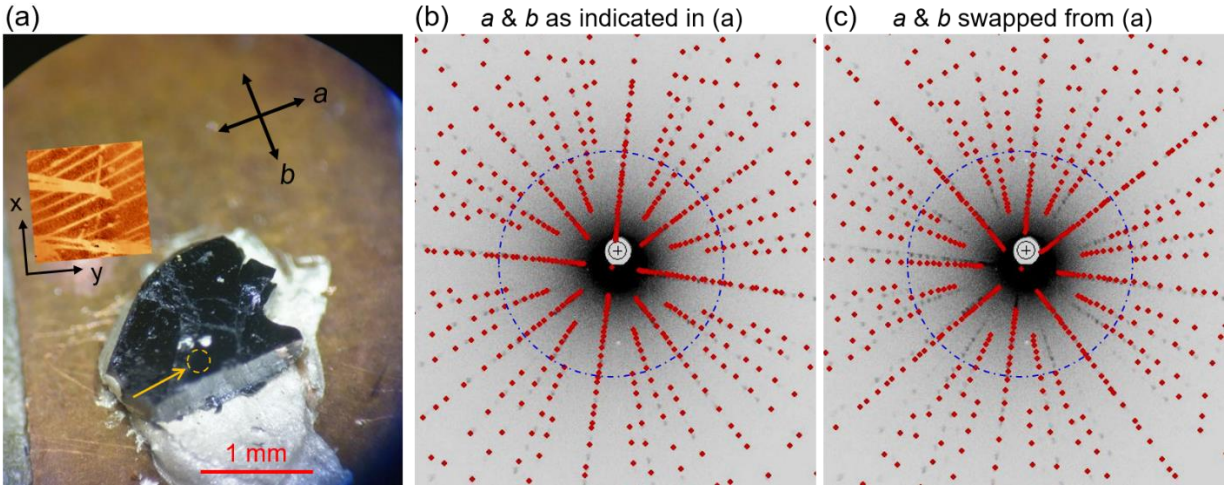

FIG. S3. (a) Picture of the Ti-doped $Ca_3Ru_2O_7$ crystal mounted on the MIM sample holder. The scan directions (*x*-axis parallel to the copper strip) are labeled in the photo. Note that the *x* and *y* axes of the MIM image in the inset are rotated from that in the main text. The scanned area is roughly inside the orange dashed circle indicated by the arrow. (b) Laue X-ray diffraction pattern (black spots) overlaid on the calculated pattern (red dots) using the lattice constants of $Ca_3(Ru_{0.9}Ti_{0.1})_2O_7$, where the *a* and *b* axes are oriented along the directions labeled in (a). (c) Same as (b) except that *a* and *b* axes are swapped for the calculated pattern (red dots). The results inside the blue dash-dotted circles show much better agreement between the measured and simulated patterns in (b) than that in (c).

In order to determine the orientation of the stripe phase, we performed careful Laue X-ray diffraction (XRD) experiment and analysis. Fig. S3a shows the $Ca_3(Ru_{0.9}Ti_{0.1})_2O_7$ crystal mounted on the MIM sample holder, where the slow scan direction (*x*) is parallel to the copper strip. We then took the crystal for high-resolution XRD measurement and determined the two principle in-plane axes. Note that the in-plane lattice constants of $Ca_3(Ru_{0.9}Ti_{0.1})_2O_7$ differ by ~ 2% between the two axes [S4]. As shown in Fig. S3b, the calculated XRD pattern (red dots) matches nicely to the experimental data (black dots) if we orient the *a* and *b* axes according to the labels in Fig. S3a, whereas swapping the assignment for these two axes in Fig. S3c leads to poor fit to the measured results. The good fitting to the XRD data in Fig. S3b also indicates that the crystal does not contain a lot of twinned domains within the X-ray spot size of ~ 1 mm. We can therefore safely assume that the crystalline orientation in the MIM images follows that in Fig. S3a.

## S4. Position of stripes at different cool down.

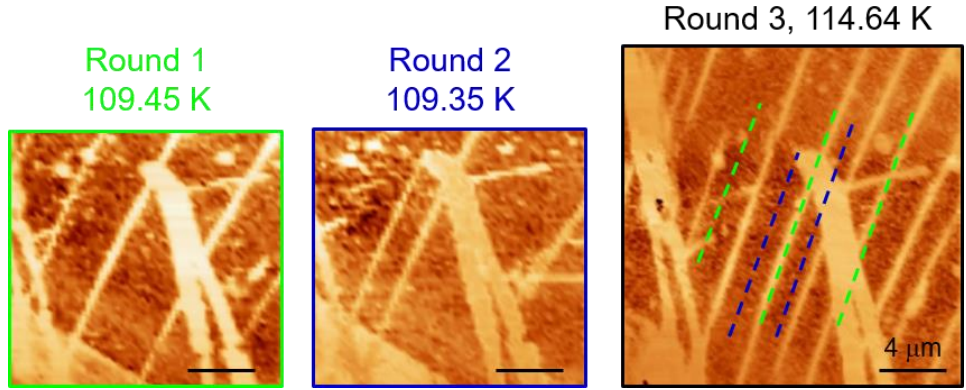

FIG. S4. AC_MIM-Re images around the same area for three different rounds of experiment. The stripes observed in the first two rounds are indicated in the last image by green and blue dashed lines.

The metallic stripe domains emerged during the Mott transition are not always pinned to specific locations. Fig. S4 shows the AC_MIM-Re images around the same area at three different rounds of cool-down. We overlay the stripes observed in the first two images (green and blue dashed lines) onto the last image, which is the 114.64 K data in Fig. 2d of the main text. It is obvious that some stripes appear at different locations from cool-down to cool-down, suggesting that such competing state emerges dynamically rather than being pinned by local disorders.

The data in Fig. S4 also illustrate that different thermal history can lead to different apparent $T_{MIT}$ in our experiment. For the Round 3 data, the sample was cooled to ~ 10 K and slowly warmed up through the transition. On the other hand, when we cooled the sample to 100 K and rapidly warmed it to 109 K before slow scans, the stripes started to appear at lower temperatures, i.e., 109.45 K in Round 1 and 109.35 K in Round 2. Note that these images are comparable to the 114.03 K data in Fig. 2d. As a result, a single $T_{MIT}$ does not apply for this complex system and the transition strongly depends on the thermal/field history.

## S5. Field-dependent MIM data at a slightly higher temperature

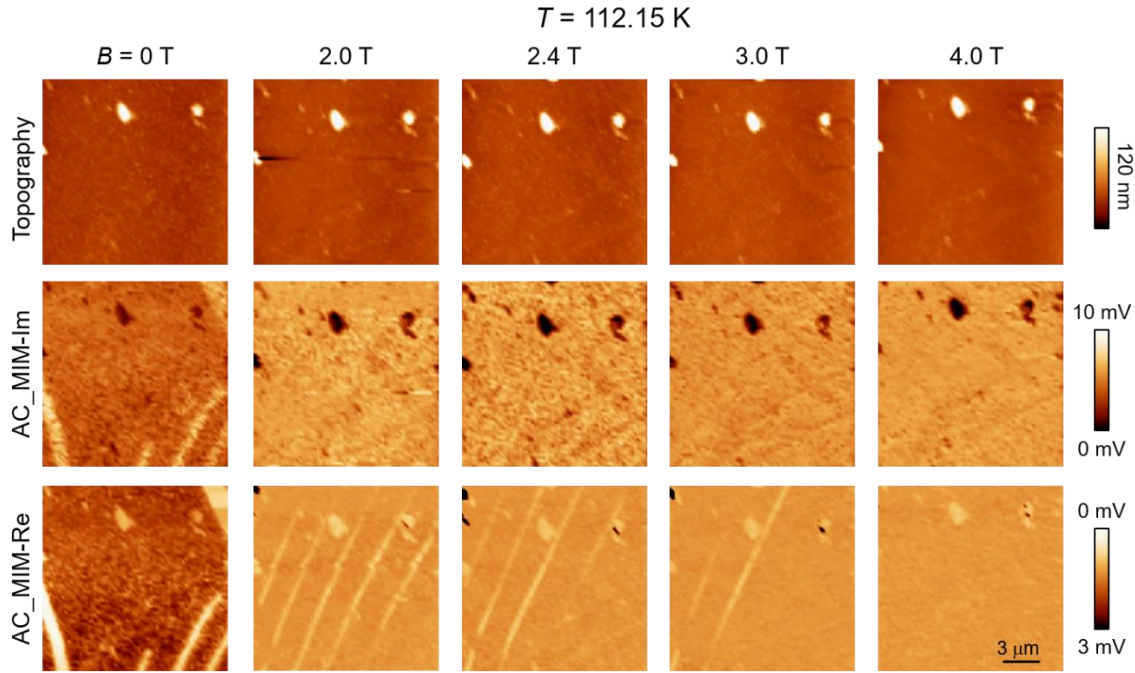

FIG. S5. Complete set of $B$-dependent images at 112.15 K. All images are 15 μm × 15 μm.

The last MIM image in Fig. 3c of the main text was not taken at the same temperature as others since the magnet was on the verge of quenching. Instead, we took another set of images at a slightly temperature of 112.15 K than that in Fig. 3c (111.40 K). At this temperature, the zero-field image was comparable to the 6.4 T data at 111.40 K. As a result, we were able to use lower fields to observe the evolution of stripes (appearing at low fields and eventually disappearing at high fields) within the PM-M phase, as displayed in Fig. S5. Other features, such as the surface particles in topographic and AC_MIM-Im images, are similar to the data in the main text.

The fact that the apparent $T_{MIT}$ differs by a few Kelvins in our data (Figs. 1 – 3 and Figs. S2 and S5) is worth some discussions. For MITs in strongly correlated materials, it is not uncommon that the transition depends on the detailed thermal or magnetic-field history of the sample. In particular, the images in Fig. 2 were taken when the sample was cooled to the base temperature of ~ 10 K and then gradually warmed up to cross the phase transition. For the data in Fig. 3, the sample was only cooled to slightly below 100 K before warming back towards the transition. The fact that we ramped up and down the magnetic fields when taking Fig. 3 and Fig. S5 may have also played a role. As a result, it is not surprising that the apparent $T_{MIT}$ differs by a few Kelvins (113 ~ 114 K in Fig. 1, 114 ~ 115 K in Fig. 2, 112 ~ 113 K in Fig. 3) in our experiment.

## S6. Stripes near the physical boundary between G-AFM-I and PM-M phases

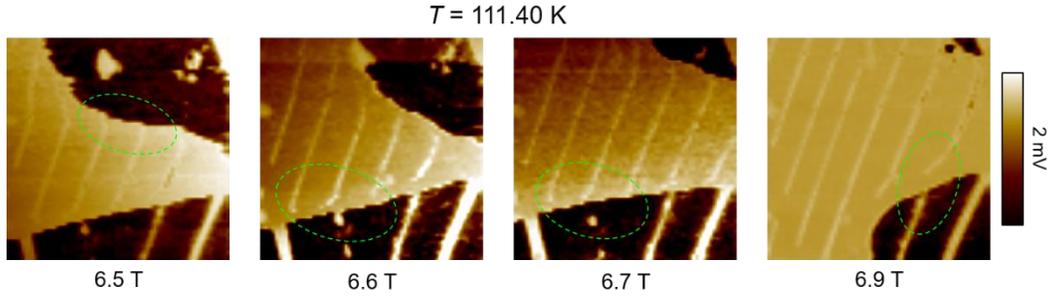

FIG. S6. *B*-dependent MIM images using a different false-color scale. All images are 15 μm × 15 μm.

Some MIM images in Fig. 3c of the main text are replotted in Fig. S6 with a different false-color scale such that the stripes are more visible near the physical boundary between G-AFM-I and PM-M phases. Several key features are observed inside the green ellipses in each image. First, the metallic stripes are not confined to the interface between the two global phases, nor are they parallel or perpendicular to the interface. The situation is different from that in the current-driven Mott transition in $Ca_2RuO_4$, where the stripes appear near the physical boundary between the two global phases [S5]. Secondly, the stripes are not continuous across the boundary, e.g., they simply terminate within the PM-M domain in the 6.5 T and 6.7 T images. Finally, for $B$ = 6.9 T, the stripe in the PM-M domain actually bends itself to avoid the connection with the stripe in the G-AFM-I domain. Future theoretical work that takes into account the elastic strain energy may be needed to explain such phenomena.